\documentclass{llncs}

\usepackage{makeidx} 
\usepackage{graphicx} 
\usepackage{subfigure} 

\usepackage{algorithm}
\usepackage{algorithmic}

\usepackage{amsmath}
\usepackage{amssymb}
\usepackage{paralist}

\newcommand{\knowref}[2][]{\ifx&#1&({\it k.\ref{#2}})\else({\it k.\ref{#1}-\ref{#2}})\fi}
\newcommand{\capref}[2][]{\ifx&#1&({\it c.\ref{#2}})\else({\it c.\ref{#1}-\ref{#2}})\fi}

\newcommand{\ie}{\emph{i.e.}}
\newcommand{\eg}{\emph{e.g.}}
\newcommand{\etal}{\emph{et al.}}

\newcommand{\tsum}{\textstyle\sum}
\newcommand{\nth}[2]{\ensuremath{{#1}^{\mbox{\scriptsize #2}}}} 

\begin{document}
\title{Evasion attacks against machine learning\\ at test time}
\author{Battista Biggio\inst{1} \and Igino Corona\inst{1} \and
Davide Maiorca\inst{1} \and Blaine Nelson\inst{2} 
 \and Nedim \v{S}rndi\'{c}\inst{3} \and Pavel Laskov\inst{3} \and
Giorgio Giacinto\inst{1} \and Fabio Roli\inst{1}}
\authorrunning{Battista Biggio \etal} 
\institute{
Dept. of Electrical and Electronic Engineering,
University of Cagliari, \\
Piazza d'Armi, 09123 Cagliari, Italy \\
\email{\{battista.biggio, igino.corona, davide.maiorca, giacinto, roli\}@diee.unica.it},\\
WWW home page: \texttt{http://prag.diee.unica.it/}
\and
Institut f\"ur Informatik, Universit\"at Potsdam, \\
August-Bebel-Stra{\ss}e 89, 14482 Potsdam, Germany \\
\email{bnelson@cs.uni-potsdam.de}
\and
Wilhelm Schickard Institute for Computer Science,
University of T\"ubingen, \\
Sand 1, 72076 T\"ubingen, Germany \\
\email{\{nedim.srndic, pavel.laskov\}@uni-tuebingen.de}
}

\maketitle              

\begin{abstract}
  In security-sensitive applications, the success of machine learning
  depends on a thorough vetting of their resistance to
  adversarial data. In one pertinent, well-motivated attack scenario,
  an adversary may attempt to evade a deployed system at test time
  by carefully manipulating attack samples.
  In this work, we present a
  simple but effective gradient-based approach that can be exploited
  to systematically assess the security of several, widely-used
  classification algorithms against evasion attacks.  Following a
  recently proposed framework for security evaluation, we
  simulate attack scenarios that exhibit different risk levels for the classifier
  by increasing the attacker's knowledge of the system and her
  ability to manipulate attack samples.
  This gives the classifier designer a better picture of the classifier performance
  under evasion attacks, and allows him to perform a more informed
  model selection (or parameter setting).
  We evaluate our approach on the relevant security task of malware
  detection in PDF files, and show that such systems can be easily evaded.
  We also sketch some countermeasures suggested by our analysis.  
  

\end{abstract}

\section{Introduction}
\label{introduction}

Machine learning is being increasingly used in security-sensitive
applications such as spam filtering, malware detection, and network
intrusion detection
\cite{biggio-IJMLC10,biggio13-tkde,dalvi04,fogla06,huang11,kloft10,kolcz09,lowd05,nelson08}.
Due to their intrinsic adversarial nature, these applications differ
from the classical machine learning setting in which the underlying
data distribution is assumed to be \emph{stationary}.  To the
contrary, in security-sensitive applications, samples (and, thus,
their distribution) can be actively manipulated by an intelligent,
adaptive adversary to confound learning; \eg, to avoid
detection, spam emails are often modified by obfuscating common spam
words or inserting words associated with legitimate
emails~\cite{biggio-IJMLC10,dalvi04,kolcz09,lowd05}.  This has led to
an arms race between the designers of learning systems and their
adversaries, which is evidenced by the increasing complexity of
modern attacks and countermeasures.  For these reasons, classical
performance evaluation techniques are not suitable to reliably assess
the security of learning algorithms, \ie, the performance degradation
caused by carefully crafted attacks~\cite{biggio13-tkde}.

To better understand the security properties of machine learning
systems in adversarial settings, paradigms from security
engineering and cryptography have been adapted to the machine
learning field~\cite{barreno-ASIACCS06,biggio13-tkde,huang11}.
Following common
security protocols, the learning system designer should use \emph{proactive} protection mechanisms that
anticipate
and prevent the adversarial impact.
This requires (\emph{i})
finding potential vulnerabilities of learning \emph{before} they are
exploited by the adversary; (\emph{ii}) investigating the
impact of the corresponding attacks (\ie, evaluating classifier
security); and (\emph{iii}) devising appropriate countermeasures if an attack is
found to significantly degrade the classifier's performance.

Two approaches have previously addressed security
issues in learning.
The min-max approach assumes the learner and attacker's loss functions are antagonistic, which yields relatively simple
optimization problems~\cite{dekel10,globerson-ICML06}. 
A more general game-theoretic approach applies
for non-antagonistic losses;
\eg, a spam filter wants to accurately identify legitimate email
while a spammer seeks to boost his spam's appeal.
Under certain conditions, such problems can be solved using a Nash equilibrium
approach~\cite{BruSch11,brueckner12}.
Both approaches provide a
\emph{secure} counterpart to their respective learning problems; \ie, an optimal anticipatory classifier.

Realistic constraints, however, are too complex and multi-faceted to be incorporated into existing game-theoretic approaches. Instead, we investigate the vulnerabilities of classification algorithms by deriving \emph{evasion attacks} in which the adversary aims to avoid detection by manipulating malicious test samples.\footnote{Note that other kinds of attacks are possible, \eg, if the adversary can manipulate the training data. A comprehensive taxonomy of attacks can be found in \cite{barreno-ASIACCS06,huang11}.}
We systematically assess classifier security in attack scenarios that exhibit increasing risk levels, simulated by increasing the attacker's knowledge of the system and her ability to manipulate attack samples.
Our analysis allows a classifier designer to understand how the
classification performance of each considered model degrades under attack,
and thus, to make more informed design choices.

The problem of evasion at test time was addressed in
prior work, but limited to linear
and convex-inducing
classifiers~\cite{dalvi04,lowd05,nelson12-jmlr}. 
In contrast, the methods presented
in Sections~\ref{sect:optimal-evasion} and \ref{sect:gradient-descent}
can generally evade linear or non-linear classifiers using a gradient-descent approach inspired by
Golland's discriminative directions technique~\cite{Gol02}.
Although we focus our analysis on widely-used classifiers such as Support Vector Machines (SVMs) and neural networks, our approach is applicable to any classifier with a differentiable discriminant function.

This paper is organized as follows.  We present
the evasion problem in Section~\ref{sect:optimal-evasion} and
our gradient-descent approach in Section~\ref{sect:gradient-descent}.  
In Section~\ref{sect:experiments} we first
visually demonstrate our attack on the task of
handwritten digit recognition, and then show its effectiveness
on a realistic application related to the detection of PDF malware.  
Finally in Section~\ref{sect:conclusions}, we summarize our
contributions, discuss possibilities for improving security,
and suggest future extensions of this work.

\section{Optimal evasion at test time}
\label{sect:optimal-evasion}

\newcommand{\classifierLBL}[1][]{\ensuremath{y^c_{#1}}}

We consider a classification algorithm $f : \mathcal X \mapsto \mathcal Y$ that assigns samples represented in some feature space $\mathbf x \in \mathcal X$ to a label in the set of predefined classes $y \in \mathcal Y = \{-1,+1\}$, where $-1$ ($+1$) represents the legitimate (malicious) class.
The classifier $f$ is trained on a dataset $\mathcal D = \{\mathbf x_{i}, y_{i}\}_{i=1}^{n}$ sampled from an underlying distribution $p(\mathbf X,Y)$.
The label $\classifierLBL = f(\mathbf x)$ given by a classifier is typically obtained by thresholding a continuous discriminant function $g : \mathcal X \mapsto \mathbb{R}$. In the sequel, we use $\classifierLBL$ to refer to the label assigned by the classifier as opposed to the true label $y$. We further assume that $f(\mathbf x) = -1$ if $g(\mathbf x) < 0$, and $+1$ otherwise.

\subsection{Adversary model}\label{subsect:adv-model}

To motivate the optimal attack strategy for evasion, it is necessary
to disclose one's assumptions of the adversary's knowledge and
ability to manipulate the data.  To this end, we exploit a general
model of the adversary that elucidates specific assumptions
about adversary's goal, knowledge of the system, and capability to
modify the underlying data distribution. The considered model is part
of a more general framework investigated in our recent work~\cite{biggio13-tkde},
which subsumes evasion and other attack scenarios. This model can
incorporate application-specific constraints in the definition of the
adversary's capability, and can thus be exploited to derive practical
guidelines for developing the optimal attack strategy.

\medskip
\noindent \textbf{Adversary's goal}.  As suggested by Laskov and Kloft \cite{laskov09}, the
adversary's goal should be defined in terms of a utility (loss)
function that the adversary seeks to maximize (minimize). In the
evasion setting, the attacker's goal is to manipulate a single (without
loss of generality, positive) sample that should be
misclassified.
Strictly speaking, it would suffice
to find a sample $\mathbf x$ such that $g(\mathbf x) < -\epsilon$ for any
$\epsilon > 0$; \ie, the attack sample only just crosses
the decision boundary.\footnote{This is also the setting adopted in
  previous work \cite{dalvi04,lowd05,nelson12-jmlr}.}  Such attacks,
however, are easily thwarted by slightly adjusting the decision threshold.  A better strategy for an attacker
would thus be to create a sample that is misclassified with high
confidence; \ie, a sample minimizing the value of the classifier's
discriminant function, $g(\mathbf x)$, subject to some feasibility constraints.

\medskip
\noindent \textbf{Adversary's knowledge}. The adversary's knowledge about her targeted
learning system may vary significantly. Such
knowledge may include:
\begin{itemize}
\item \label{know-1} the training set or part of it;
\item\label{know-2} the feature representation of each sample;
  \ie, how \emph{real} objects such as emails, network packets 
  are mapped into the classifier's feature space; 
\item\label{know-3} the type of a learning algorithm and the form of
  its decision function;
\item\label{know-4} the (trained) classifier model; \eg, weights of a linear classifier; 
\item\label{know-5} or feedback from the classifier; \eg, classifier labels for samples chosen by the adversary.
\end{itemize}

\medskip
\noindent \textbf{Adversary's capability}. In the evasion scenario, the
adversary's capability is limited to modifications of test data; \ie altering
the training data is not allowed. However, under this restriction, variations in attacker's power may include:

\begin{itemize}
\item modifications to the input data (limited or unlimited);
\item modifications to the feature vectors (limited or unlimited);
\item or independent modifications to specific features (the semantics of the
  input data may dictate that certain features are interdependent).
\end{itemize}

Most of the previous work on evasion attacks assumes that the attacker can 
arbitrarily change every feature~\cite{brueckner12,dekel10,globerson-ICML06}, 
but they constrain the degree of manipulation, \eg, limiting the number of modifications, or their total cost. However, many real domains impose stricter restrictions.
For example, in the task of PDF malware detection~\cite{maiorca,Smutz,Srndic},
removal of content is not feasible, and content addition may cause correlated changes in the feature vectors.

\subsection{Attack scenarios}

In the sequel, we consider two attack scenarios characterized by
different levels of adversary's knowledge of the attacked system discussed below.

\medskip
\noindent \textbf{Perfect knowledge (PK).} In this setting, we assume that the adversary's goal is to minimize
$g(\mathbf x)$, and that she has perfect knowledge of the targeted
classifier; \ie, the adversary knows the feature space, the type of the classifier,
and the trained model. The adversary can transform attack points in the test data but must remain within a maximum distance of $d_{\rm max}$ from the original attack sample.
We use $d_{\rm max}$ as parameter in our evaluation to simulate
increasingly pessimistic attack scenarios by giving the adversary
greater freedom to alter the data.

The choice of a suitable distance measure $d : \mathcal X \times \mathcal X \mapsto \mathbb{R}^{+}$ is application specific~\cite{dalvi04,lowd05,nelson12-jmlr}.
Such a distance measure
should reflect the adversary's effort required to manipulate samples
or the cost of these manipulations.
For example, in spam filtering, the attacker may be bounded by a certain
number of words she can manipulate, so as not to lose the semantics
of the spam message.

\medskip
\noindent \textbf{Limited knowledge (LK).} Here, we again assume that the adversary aims to minimize the
discriminant function $g(\mathbf x)$ under the same constraint that
each transformed attack point must remain within a maximum distance of $d_{\rm max}$ from the corresponding original attack sample.
We further assume that the attacker knows the feature
representation and the type of the classifier,
but does not know either the learned classifier $f$ or its training data $\mathcal D$,
and hence can not directly compute $g(\mathbf x)$.
However, we assume that she can collect a surrogate dataset $\mathcal D^{\prime} = \{\hat{\mathbf x}_{i}, \hat y_{i}\}_{i=1}^{n_{q}}$ of $n_{q}$ samples drawn from the same underlying distribution $p(\mathbf X,Y)$ from which $\mathcal D$ was drawn.
This data may be collected by an adversary in
several ways; \eg, by sniffing some network traffic during the
classifier operation, or by collecting legitimate and spam
emails from an alternate source.

Under this scenario, the adversary proceeds by approximating the
discriminant function $g(\mathbf x)$ as $\hat g(\mathbf x)$, where $\hat g(\mathbf x)$ is the discriminant function of a surrogate classifier $\hat f$ learnt on $\mathcal D^{\prime}$. The amount of the surrogate data, $n_{q}$, is an attack parameter in our experiments. 
Since the adversary wants her surrogate $\hat f$ to closely approximate the
targeted classifier $f$, it stands to reason that she should learn
$\hat f$ using the labels assigned by the targeted classifier $f$, when such feedback is available. In this case, instead of using the true class labels $\hat y_{i}$ to train $\hat f$, the adversary can query $f$ with the samples of $\mathcal D^{\prime}$ and subsequently learn using the labels $\hat{y}^{c}_{i} = f(\hat{\mathbf x}_{i})$ for each ${\mathbf x}_{i}$.

\subsection{Attack strategy}
\label{sect:attack-strategy}

Under the above assumptions, for any target malicious sample $\mathbf x^{0}$ (the adversary's desired instance), an optimal attack strategy finds a sample $\mathbf x^{\ast}$ to minimize $g(\cdot)$ or its estimate $\hat g(\cdot)$, subject to a bound on its distance\footnote{One can also incorporate additional application-specific constraints on the attack samples. For instance, the box constraint $0 \leq x_f \leq 1$ can be imposed if the \nth{f}{th} feature is normalized in $[0,1]$, or $x^{0}_f \leq x_f$ can be used if the \nth{f}{th} feature of the target $\mathbf x^{0}$ can be only incremented.} from $\mathbf x^{0}$:
\begin{eqnarray}
\label{eq:evasion-obj} \mathbf x^{\ast} = \arg \min_{\mathbf x} & \; & \hat g(\mathbf x)  \\ 
\nonumber \rm{s.t.} & \; & d(\mathbf x,\mathbf x^{0}) \leq d_{\rm max}. 
\end{eqnarray}

\begin{figure}[htb]
\begin{center}
\includegraphics[width=0.49\textwidth]{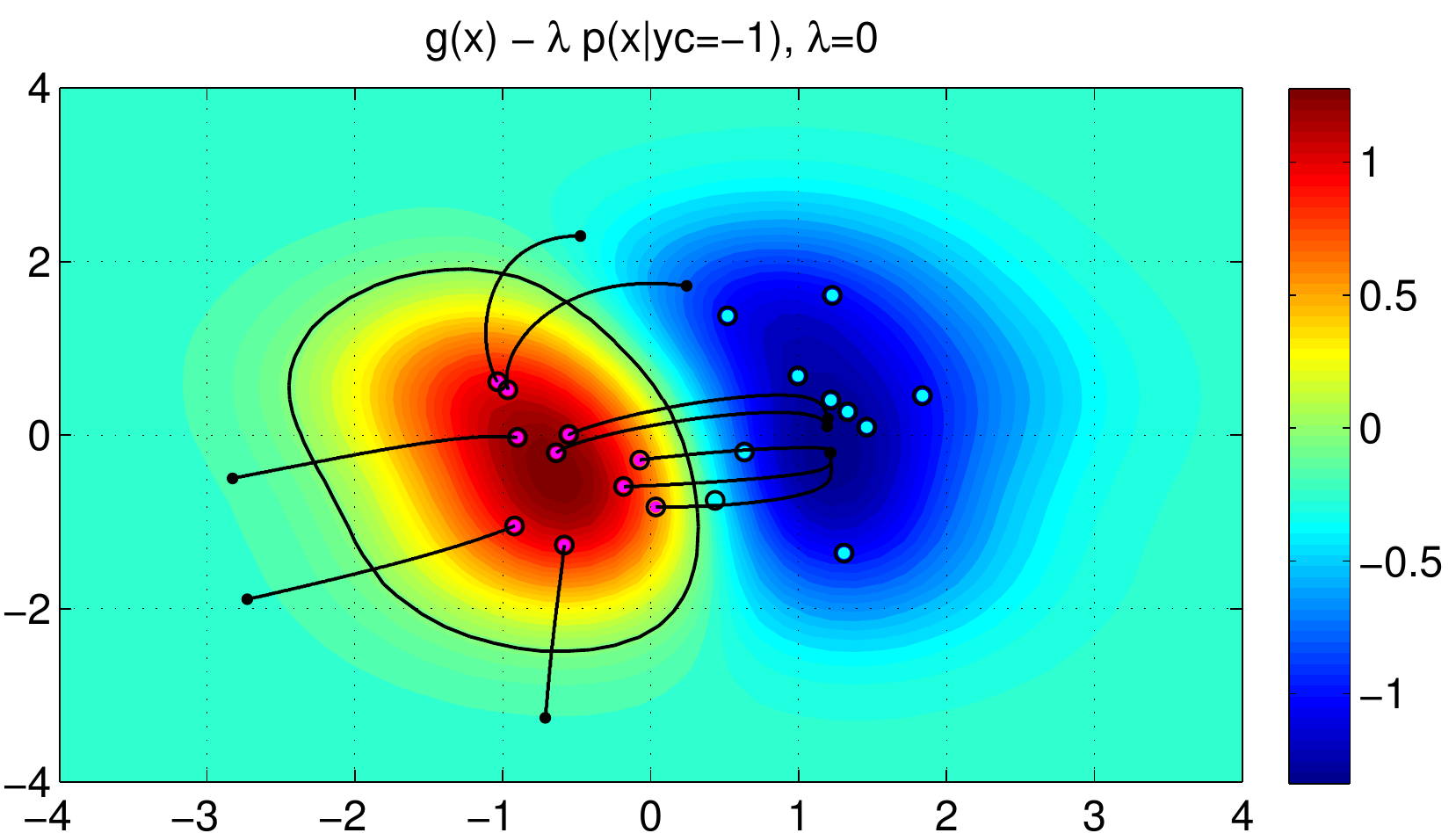}
\includegraphics[width=0.49\textwidth]{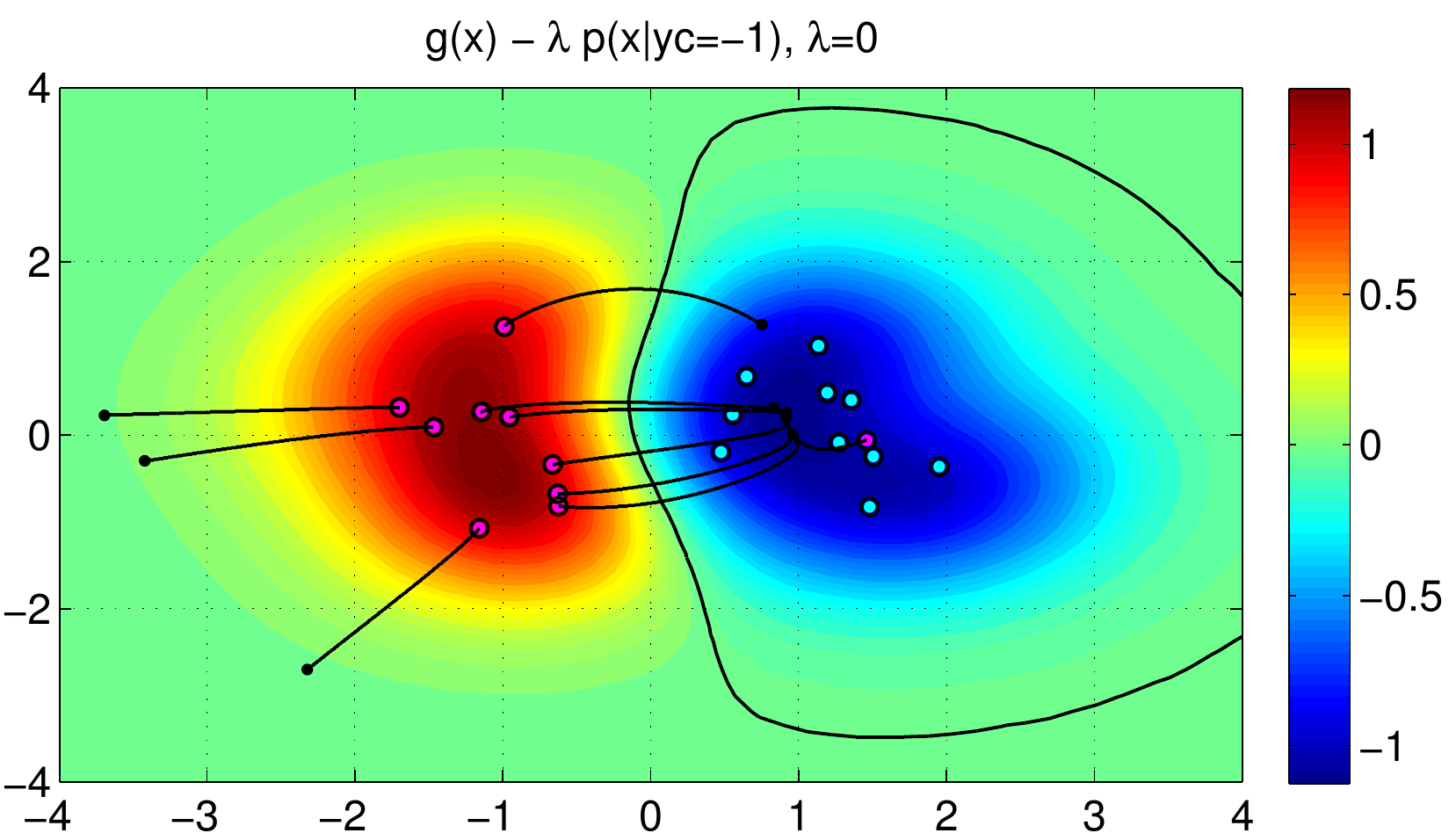}
\includegraphics[width=0.49\textwidth]{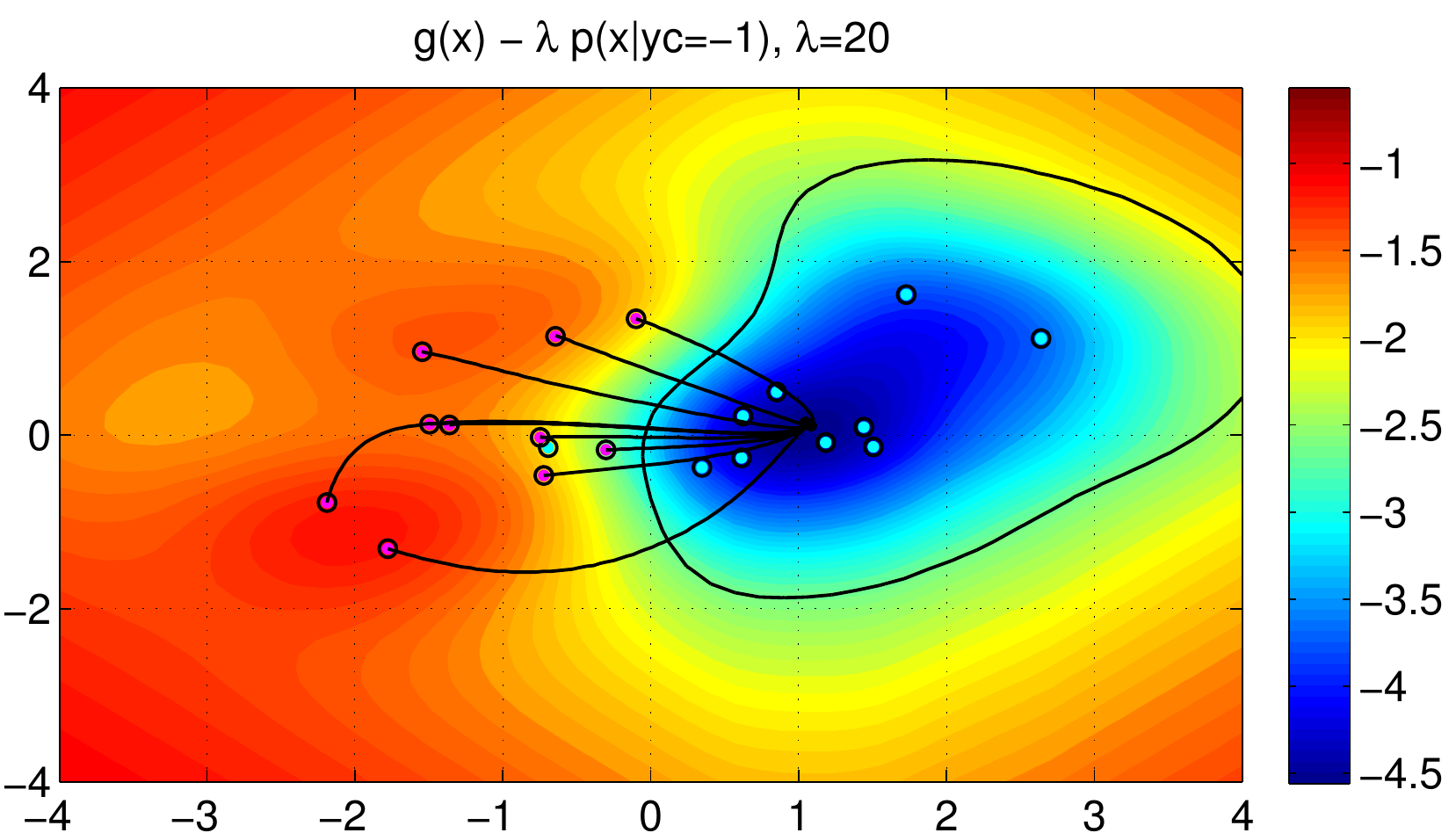}
\caption{Different scenarios for gradient-descent-based evasion procedures. In each, the function $g(\mathbf x)$ of the learned classifier is plotted with a color map with high values (red-orange-yellow) for the malicious class, and low values (green-cyan-blue) for the legitimate class. The decision boundary is shown in black. For every malicious sample, we plot the gradient descent path against a classifier with a closed boundary around the malicious class (\textbf{top-left}) and against a classifier with a closed boundary around the benign class (\textbf{top-right}). Finally, we plot the modified objective function of Eq.~\eqref{eq:obj-function} and the resulting descent paths against a classifier with a closed boundary around the benign class (\textbf{bottom}).}
\label{fig:attack-strategy}
\end{center}
\end{figure}

Generally, this is a non-linear optimization problem. One may approach
it with many well-known techniques, like gradient descent, or
quadratic techniques such as Newton's method, BFGS, or L-BFGS. We
choose a gradient-descent procedure. However, $\hat g(\mathbf x)$ may
be non-convex and descent approaches may not achieve a global optima.
Instead, the descent path may lead to a flat region (local minimum)
outside of the samples' support (\ie, where $p(\mathbf x) \approx 0$)
where the attack sample may or may not evade depending on the behavior
of $g$ in this unsupported region (see left and middle plots in
Figure~\ref{fig:attack-strategy}).

Locally optimizing $\hat{g}(\mathbf x)$ with gradient descent is
particularly susceptible to failure due to the nature of a discriminant function.
Besides its shape, for many classifiers, $g(\mathbf x)$ is equivalent
to a posterior estimate $p(\classifierLBL=-1 | \mathbf x)$; \eg, for
neural networks, and SVMs \cite{platt99}.
The discriminant function does not incorporate the evidence we have
about the data distribution, $p(\mathbf x)$, and thus, using gradient
descent to optimize Eq.~\ref{eq:evasion-obj} may lead into unsupported
regions ($p(\mathbf x) \approx 0$).
Because of the insignificance of these regions, the value of $g$ is relatively unconstrained by criteria such as risk minimization.
This problem is compounded by our finite (and possibly small) training set, since it provides little evidence in these regions to constrain the shape of $g$.
Thus, when our gradient descent procedure produces an evasion example
in these regions, the attacker cannot be confident that this sample
will actually evade the corresponding classifier. 
Therefore, to increase the probability of successful evasion, the
attacker should favor attack points from densely populated regions of
legitimate points, where the estimate $\hat g(\mathbf x)$ is more reliable
(closer to the real $g(\mathbf x)$), and tends to become negative in value.

To overcome this shortcoming, we introduce an additional component into our attack objective, which estimates $p(\mathbf x | \classifierLBL=-1)$ using a density estimator. This term acts as a penalizer for $\mathbf x$ in low density regions and is weighted by a parameter $\lambda \ge 0$ yielding the following modified optimization problem:
\begin{eqnarray}
\label{eq:obj-function}
\arg \min_x & &  F(\mathbf x) = \hat g(\mathbf x) - \frac{\lambda}{n} \sum_{i | \classifierLBL[i] = -1} k \left( \tfrac{\mathbf x-\mathbf x_{i}}{h} \right) \\
\label{eq:constraint}
\rm{s.t.} & & d(\mathbf x,\mathbf x^{0}) \leq d_{\rm max} \enspace,
\end{eqnarray}
where $h$ is a bandwidth parameter for a kernel density estimator (KDE), and $n$ is the number of benign samples ($\classifierLBL=-1$) available to the adversary. 
This alternate objective trades off between minimizing $\hat g(\mathbf x)$ (or $p(\classifierLBL=-1 | \mathbf x)$) and maximizing the estimated density $p(\mathbf x | \classifierLBL=-1)$. 
The extra component favors attack points that imitate features of known legitimate samples.
In doing so, it reshapes the objective function and thereby biases the resulting gradient descent towards regions where the negative class is concentrated (see the bottom plot in Fig.~\ref{fig:attack-strategy}). 
This produces a similar effect to that shown by \emph{mimicry} attacks in network intrusion detection~\cite{fogla06}.\footnote{Mimicry attacks \cite{fogla06} consist of camouflaging malicious network packets to evade anomaly-based intrusion detection systems by mimicking the characteristics of the legitimate traffic distribution.}
For this reason, although our setting is rather different, in the sequel we refer to this extra term as the \emph{mimicry} component.

Finally, we point out that, when mimicry is used ($\lambda > 0$), our gradient descent clearly follows a suboptimal path compared to the case when only $g(\mathbf x)$ is minimized ($\lambda = 0$). Therefore, more modifications may be required to reach the same value of $g(\mathbf x)$ attained when $\lambda=0$. However, as previously discussed, when $\lambda=0$, our descent approach may terminate at a local minimum where $g(\mathbf x)>0$, without successfully evading detection. This behavior can thus be qualitatively regarded as a trade-off between the probability of evading the targeted classifier and the number of times that the adversary must modify her samples.

\section{Gradient descent attacks}
\label{sect:gradient-descent}

Algorithm~\ref{lab:attack} solves the optimization problem in Eq.~\ref{eq:obj-function} via gradient descent. We assume $g(\mathbf x)$ to be differentiable almost everywhere (subgradients may be used at discontinuities).
However, note that if $g$ is non-differentiable or insufficiently smooth,
one may still use the mimicry / KDE term of Eq.~\eqref{eq:obj-function} as a search heuristic. This investigation is left to future work.  

\begin{algorithm}[t]
  \caption{Gradient-descent evasion attack}
  \textbf{Input:} $\mathbf x^{0}$, the initial attack point; $t$, the step size; $\lambda$, the trade-off parameter; $\epsilon > 0$ a small constant.\\
  \textbf{Output:} $\mathbf x^{*}$, the final attack point.
  \begin{algorithmic}[1]
    \STATE{$m \gets 0$.}
    \REPEAT
          \STATE{$m \gets m + 1$}
      \STATE{Set $\nabla F(\mathbf x^{m-1})$ to a unit vector aligned
      with $\nabla g(\mathbf x^{m-1}) - \lambda \nabla{p(\mathbf x^{m-1} | \classifierLBL=-1)}$.} 
     \STATE{$\mathbf x^{m} \leftarrow \mathbf x^{m-1} - t \nabla F(\mathbf x^{m-1})$}
     \IF{$d(\mathbf x^{m},\mathbf x^{0}) > d_{\rm max}$}
     \STATE{Project $\mathbf x^{m}$ onto the boundary of the feasible region.}
     \ENDIF
      \UNTIL{$F\left(\mathbf x^{m}\right) - F\left( \mathbf x^{m-1}\right) < \epsilon$}
      \STATE{\textbf{return:} $\mathbf x^{*} = \mathbf x^{m}$}
  \end{algorithmic}\label{lab:attack}
\end{algorithm}

\subsection{Gradients of discriminant functions}
\label{sect:gradient-disc}

\bigskip
\noindent \textbf{Linear classifiers.} Linear discriminant functions are $g(\mathbf x) = \langle \mathbf w, \mathbf x \rangle + b$ where $\mathbf w \in \mathbb{R}^{d}$ is the feature weights and $b \in \mathbb{R}$ is the bias. Its gradient is $\nabla g(\mathbf x) = \mathbf w$.

\bigskip
\noindent \textbf{Support vector machines.} For SVMs, $g(\mathbf x) = \sum_i \alpha_i y_i k(\mathbf x,\mathbf x_{i}) + b$. The gradient is thus $\nabla g(\mathbf x) = \sum_i \alpha_i y_i \nabla k(\mathbf x,\mathbf x_{i})$. In this case, the feasibility of our
approach depends on whether the kernel gradient $\nabla k(\mathbf x,\mathbf x_{i})$ is computable as it is for many numeric kernels. For instance, the gradient of the RBF kernel, $k(\mathbf x,\mathbf x_{i}) = \exp\{-\gamma \|\mathbf x-\mathbf x_{i}\|^2\}$, is $\nabla k(\mathbf x,\mathbf x_{i}) = -2 \gamma \exp\{-\gamma \|\mathbf x-\mathbf x_{i}\|^2\}(\mathbf x-\mathbf x_{i})$, and for the polynomial kernel, $k(\mathbf x,\mathbf x_{i}) = (\langle \mathbf x, \mathbf x_{i} \rangle + c)^{p}$, it is $\nabla k(\mathbf x,\mathbf x_{i}) = p(\langle \mathbf x, \mathbf x_{i} \rangle + c)^{p-1}\mathbf x_{i}$.

\begin{figure}[t]
\begin{center}
\includegraphics[width=0.4\textwidth]{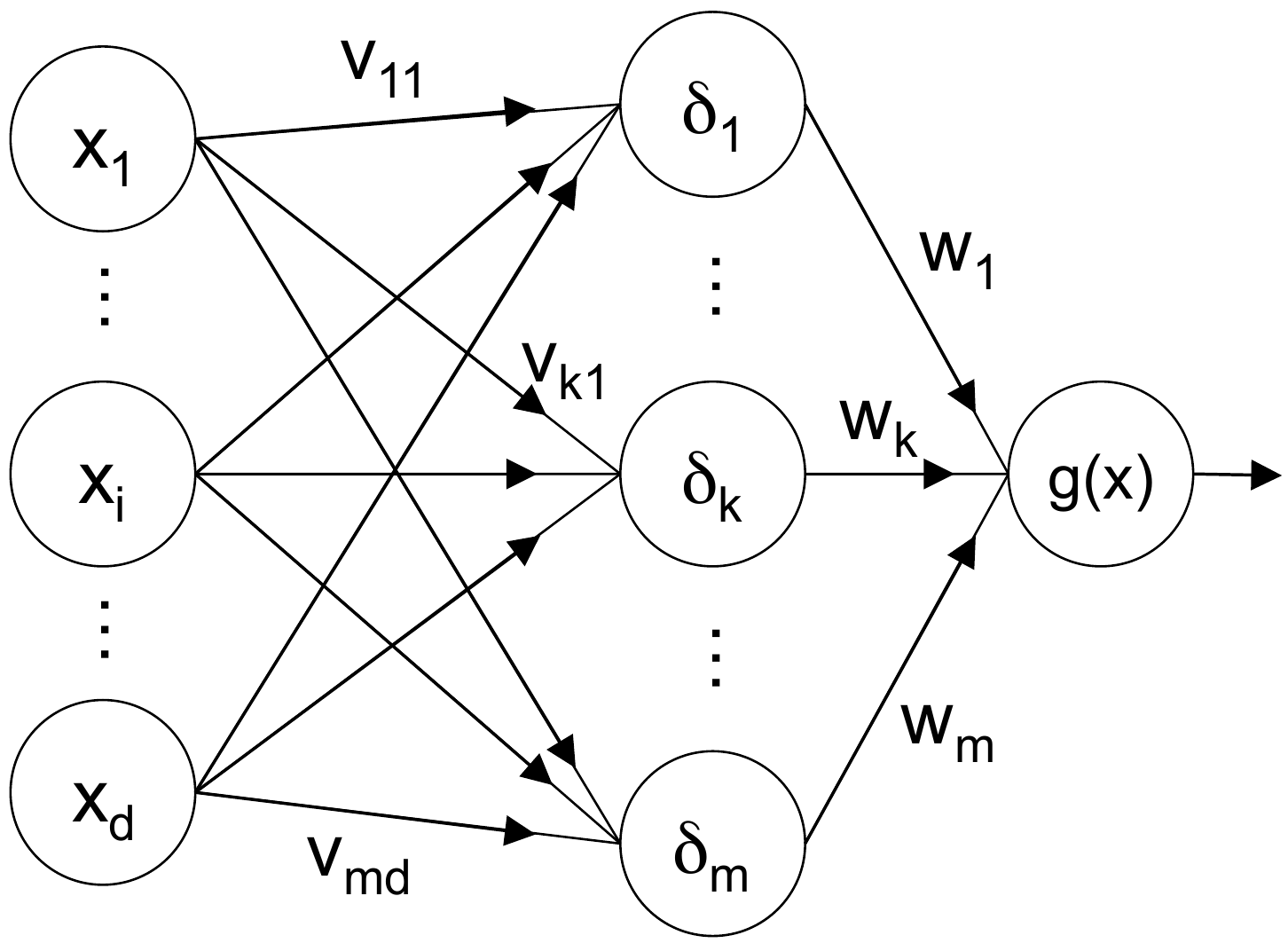}
\caption{The architecture of a multi-layer perceptron with a single hidden layer.}
\label{fig:nnet}
\end{center}
\end{figure}

\bigskip
\noindent \textbf{Neural networks.} For a multi-layer perceptron with a single hidden layer of $m$ neurons and a sigmoidal activation function, we decompose its discriminant function $g$ as follows (see Fig.~\ref{fig:nnet}): 
 $g(\mathbf x) = (1+e^{-h(\mathbf x)})^{-1}$, $h(\mathbf x) = \sum_{k=1}^{m} w_{k} \delta_{k}(\mathbf x) + b$, $\delta_{k}(\mathbf x) = (1+e^{-h_{k}(\mathbf x)})^{-1}$,
 $h_{k}(\mathbf x) = \sum_{j=1}^{d} v_{kj} x_{j} + b_{k}$. From the chain rule, the $i^\textrm{th}$ component of $\nabla g(\mathbf x)$ is thus given by:
\[ 
\tfrac{\partial g}{\partial x_{i}} = \tfrac{\partial g}{\partial h} \tsum_{k=1}^{m} \tfrac{\partial h}{\partial \delta_{k}}  \tfrac{\partial \delta_{k}}{\partial h_{k}} \tfrac{\partial h_{k}}{\partial x_{i}}=
g(\mathbf x)(1- g(\mathbf x)) \tsum_{k=1}^{m} w_{k} \delta_{k}(\mathbf x) (1-\delta_{k}(\mathbf x)) v_{ki}
\enspace.
\]

\subsection{Gradients of kernel density estimators}
\label{sect:gradients-kde}

Similarly to SVMs, the gradient of kernel density estimators depends on the kernel gradient.
We consider generalized RBF kernels of the form $k \left (\frac{\mathbf x-\mathbf x_{i}}{h} \right ) = \exp{\left (  -\frac{d(\mathbf x,\mathbf x_{i})}{h}\right )}$, where $d(\cdot,\cdot)$ is any suitable distance function.
Here we use the same distance $d(\cdot,\cdot)$ defined in Eq.~\eqref{eq:constraint},
but, in general, they can be different.
For $\ell_{2}$- and $\ell_{1}$-norms (\ie, RBF and Laplacian kernels), the KDE (sub)gradients are respectively given by:
\begin{eqnarray*}
-\tfrac{2}{n h} \tsum_{i | \classifierLBL[i] = -1}  \exp{\left (  -\tfrac{ \|\mathbf x-\mathbf x_{i}\|^{2}_2 }{h}\right )} (\mathbf x - \mathbf x_{i}) \enspace, \\
-\tfrac{1}{n h} \tsum_{i | \classifierLBL[i] = -1}  \exp{\left (  -\tfrac{ \|\mathbf x-\mathbf x_{i}\|_1}{h}\right )} (\mathbf x - \mathbf x_{i}) \enspace.
\end{eqnarray*}

Note that the scaling factor here is proportional to $O(\frac{1}{nh})$.
Therefore, to influence gradient descent with a significant mimicking effect, the value of $\lambda$ in the objective function should be chosen such that the value of $\frac{\lambda}{nh}$ is comparable with (or higher than) the range of values of the discriminant function $\hat g(\mathbf x)$.

\subsection{Descent in discrete spaces}

In discrete spaces, gradient approaches travel through
infeasible portions of the feature space. In such cases, we need to find
a feasible neighbor $\mathbf x$ that
maximally decrease $F(\mathbf x)$.  A simple approach to this problem is to
probe $F$ at every point in a small neighborhood of $\mathbf x$,
which would however require a large number of queries. For classifiers
with a differentiable decision function, we can instead select the neighbor
whose change best aligns with $\nabla F(\mathbf x)$ and decreases the objective function; \ie, to prevent overshooting a minimum.

\section{Experiments}
\label{sect:experiments}

In this section, we first report a toy example from the
MNIST handwritten digit classification task
\cite{LeCun95} to visually demonstrate how the
proposed algorithm modifies digits to mislead classification.
We then show the effectiveness of the proposed attack on a more
realistic and practical scenario: the detection of malware in PDF
files.

\subsection{A toy example on handwritten digits}
\label{sect:toy-example}

Similar to Globerson and Roweis~\cite{globerson-ICML06}, we consider
discriminating between two distinct digits from the MNIST
dataset~\cite{LeCun95}.  Each digit example is represented as a
gray-scale image of $28 \times 28$ pixels
arranged in raster-scan-order to give feature vectors of $d = 28 \times 28 = 784$ values.  We normalized each feature (pixel)
$\mathbf x \in [0,1]^{d}$ by dividing its value by $255$, and we
constrained the attack samples to this range.  Accordingly, we
optimized Eq.~\eqref{eq:obj-function}
subject to $0 \leq x_f \leq 1$ for all $f$.

\begin{figure}[t]
\begin{center}
\includegraphics[width=0.78\textwidth]{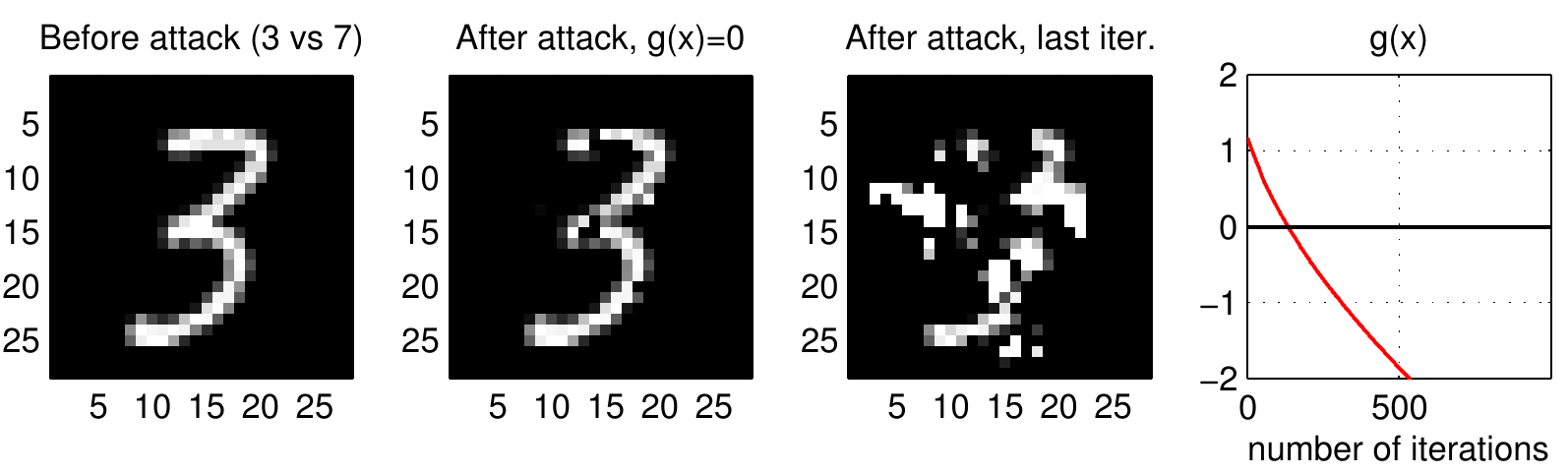}\\
\includegraphics[width=0.78\textwidth]{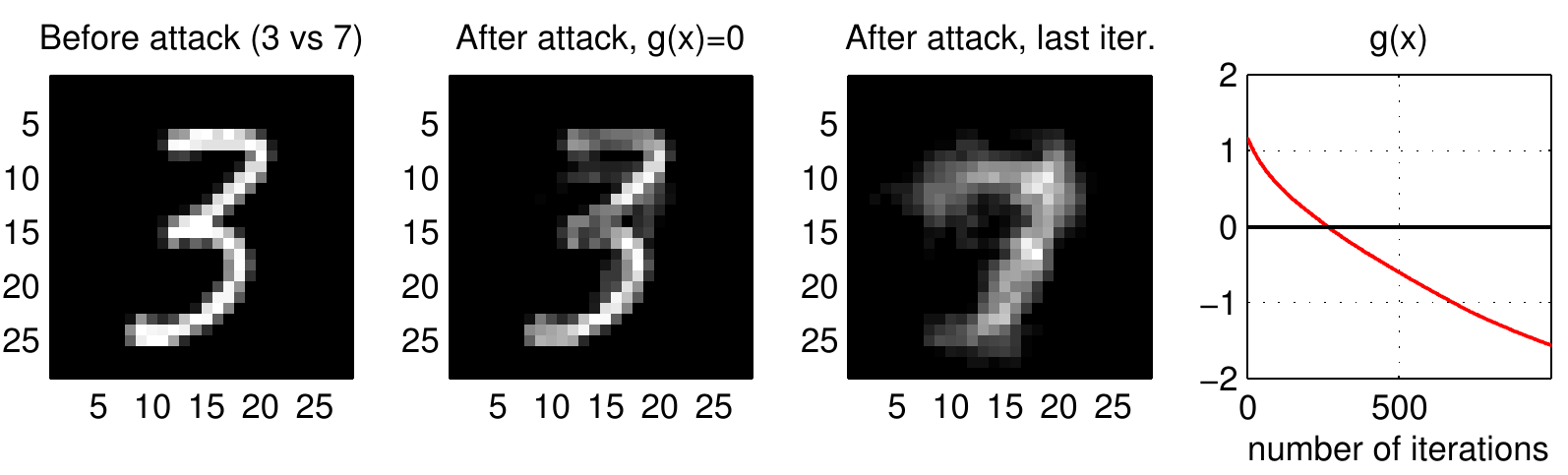}
\caption{Illustration of the gradient attack on the digit data, for $\lambda = 0$ (\textbf{top row}) and $\lambda=10$ (\textbf{bottom row}). Without a mimicry component ($\lambda = 0$) gradient descent quickly decreases $g$ but the resulting attack image does not resemble a ``7''. In contrast, the attack minimizes $g$ slower when mimicry is applied ($\lambda = 0$) but the final attack image closely resembles a mixture between ``3'' and ``7'', as the term ``mimicry'' suggests.
}
\label{fig:mnist}
\end{center}
\end{figure}

We only consider the perfect knowledge (PK) attack scenario.
We used the \emph{Manhattan} distance ($\ell_{1}$-norm), $d$, both for the kernel density
estimator (\ie, a Laplacian kernel) and for the
constraint $d(\mathbf x,\mathbf x^{0}) \leq d_{\rm max}$ in Eq.~\eqref{eq:constraint},
which bounds the total difference between the gray level
values of the original image $\mathbf x^{0}$ and the attack image $\mathbf x$. 
We used $d_{\rm max} = \frac{5000}{255}$ to limit 
the total gray-level change to $5000$. At
each iteration, we increased the $\ell_{1}$-norm value of $\mathbf x-\mathbf x^{0}$ by
$\frac{10}{255}$, or equivalently, we changed the total gray level by $10$. This is effectively the gradient step size.
The targeted classifier was an SVM with the linear kernel and $C=1$. We randomly chose $100$ training samples and applied the attacks to a correctly-classified positive sample.

In Fig.~\ref{fig:mnist} we illustrate gradient attacks in which a ``3'' is to be
misclassified as a ``7''.
The left image shows the initial attack point, the middle image shows
the first attack image misclassified as legitimate, and the right
image shows the attack point after 500 iterations.  When $\lambda=0$,
the attack images exhibit only a weak resemblance to the target class
``7'' but are, nevertheless, reliably misclassified. This is the same
effect demonstrated in the top-left plot of
Fig.~\ref{fig:attack-strategy}: the classifier is evaded by making
the attack sample sufficiently dissimilar from the malicious class.  Conversely, when $\lambda=10$, the attack images strongly resemble the target class
because the mimicry term favors samples that are more similar to the
target class. This is the same effect seen in the bottom
plot of Fig.~\ref{fig:attack-strategy}.

Finally note that, as expected, $g(\mathbf x)$ tends to decrease more gracefully when mimicry is used, as we follow a suboptimal descent path.
Since the targeted classifier can be easily evaded when $\lambda=0$, exploiting the mimicry component would not be the optimal choice in this case. However, in the case of limited knowledge, as discussed at the end of Section~\ref{sect:attack-strategy}, mimicry may allow us to trade for a higher probability of evading the targeted classifier, at the expense of a higher number of modifications.

\subsection{Malware detection in PDF files}
\label{sect:malware-pdf}

We now focus on the task of discriminating between legitimate and
malicious PDF files, a popular medium for disseminating
malware~\cite{IBM}. PDF files are excellent vectors for
malicious-code, due to their flexible \emph{logical structure}, which
can described by a hierarchy of interconnected objects. 
As a result, an attack can
be easily hidden in a PDF to circumvent file-type filtering.
The PDF format further allows a wide variety of
resources to be embedded in the document including \texttt{JavaScript},
\texttt{Flash}, and even binary programs. The type of the
embedded object is specified by \emph{keywords}, and its content is in
a \emph{data stream}.  Several recent works proposed machine-learning
techniques for detecting malicious PDFs using the file's logical
structure to accurately identify the malware~\cite{maiorca,Smutz,Srndic}.
In this case study, we use the feature representation
of Maiorca~\etal~\cite{maiorca} in which each feature corresponds to the tally of occurrences of a given keyword.

The PDF structure imposes natural constraints on attacks.  Although
it is difficult to \emph{remove} an embedded object (and
its keywords) from a PDF without corrupting the PDF's file
structure, it is rather easy to \emph{insert} new objects (and, thus,
keywords) through the addition of a new \emph{version} to the PDF
file~\cite{Refer}.  In our feature representation, this is equivalent
to allowing only feature increments, \ie, requiring $\mathbf x^{0} \leq \mathbf x$ as
an additional constraint in the optimization problem given by
Eq.~\eqref{eq:obj-function}.  Further, the total difference in keyword 
counts between two samples is their
\emph{Manhattan} distance, which we again use for the kernel
density estimator and the constraint in
Eq.~\eqref{eq:constraint}. Accordingly, $d_{\rm max}$ is the
maximum number of additional keywords that an attacker 
can add to the original $\mathbf x^{0}$.

\medskip
\noindent \textbf{Experimental setup.} For experiments, we used a PDF corpus with
500 malicious samples from the \emph{Contagio}
dataset\footnote{\url{http://contagiodump.blogspot.it}} and 500 benign
samples collected from the web. We randomly split the data into five
pairs of training and testing sets with 500 samples each to average
the final results. The features (keywords) were extracted from each
training set as described in \cite{maiorca}. On average, $100$ keywords were found in each run.
Further, we also bounded the maximum value of each feature to $100$, as this value was found to be close to the $95^{\rm th}$ percentile for each feature. This limited the influence of outlying samples.

We simulated the \emph{perfect} knowledge (PK) and the \emph{limited} knowledge (LK) scenarios described in Section~\ref{subsect:adv-model}. In the LK case, we set the number of samples used to learn the surrogate classifier to $n_{g} =100$. The reason is to demonstrate that even with a dataset as small as the 20\% of the original training set size, the adversary may be able to evade the targeted classifier with high reliability.
Further, we assumed that the adversary uses feedback from the \emph{targeted} classifier $f$; \ie, the labels $\hat y^{c}_{i} =  f(\hat{\mathbf x}_{i})$ for each surrogate sample $\hat {\mathbf x}_{i} \in \mathcal D^{\prime}$.\footnote{Similar results were also obtained using the true labels (without relabeling), since the targeted classifiers correctly classified almost all samples in the test set.}

As discussed in Section~\ref{sect:gradients-kde}, the value of $\lambda$ is chosen according to the scale of the discriminant function $g(\mathbf x)$, the bandwidth parameter $h$ of the kernel density estimator, and the number of legitimate samples $n$ in the surrogate training set.
For computational reasons, to estimate the value of the KDE at $\mathbf x$, we only consider the $50$ nearest (legitimate) training samples to $\mathbf x$; therefore, $n \leq 50$ in our case.
The bandwidth parameter was set to $h=10$, as this value provided a proper rescaling of the Manhattan distances observed in our dataset for the KDE. We thus set $\lambda = 500$ to be comparable with $O(nh)$.

For each targeted classifier and training/testing pair, we learned five surrogate classifiers by randomly selecting $n_{g}$ samples from the test set, and we averaged their results.
For SVMs, we sought a surrogate classifier that would correctly match the labels from the targeted classifier; thus, we used parameters $C=100$, and $\gamma=0.1$ (for the RBF kernel) to heavily penalize training errors.

\begin{figure}[htbp]
\begin{center}
\includegraphics[width=0.495\textwidth]{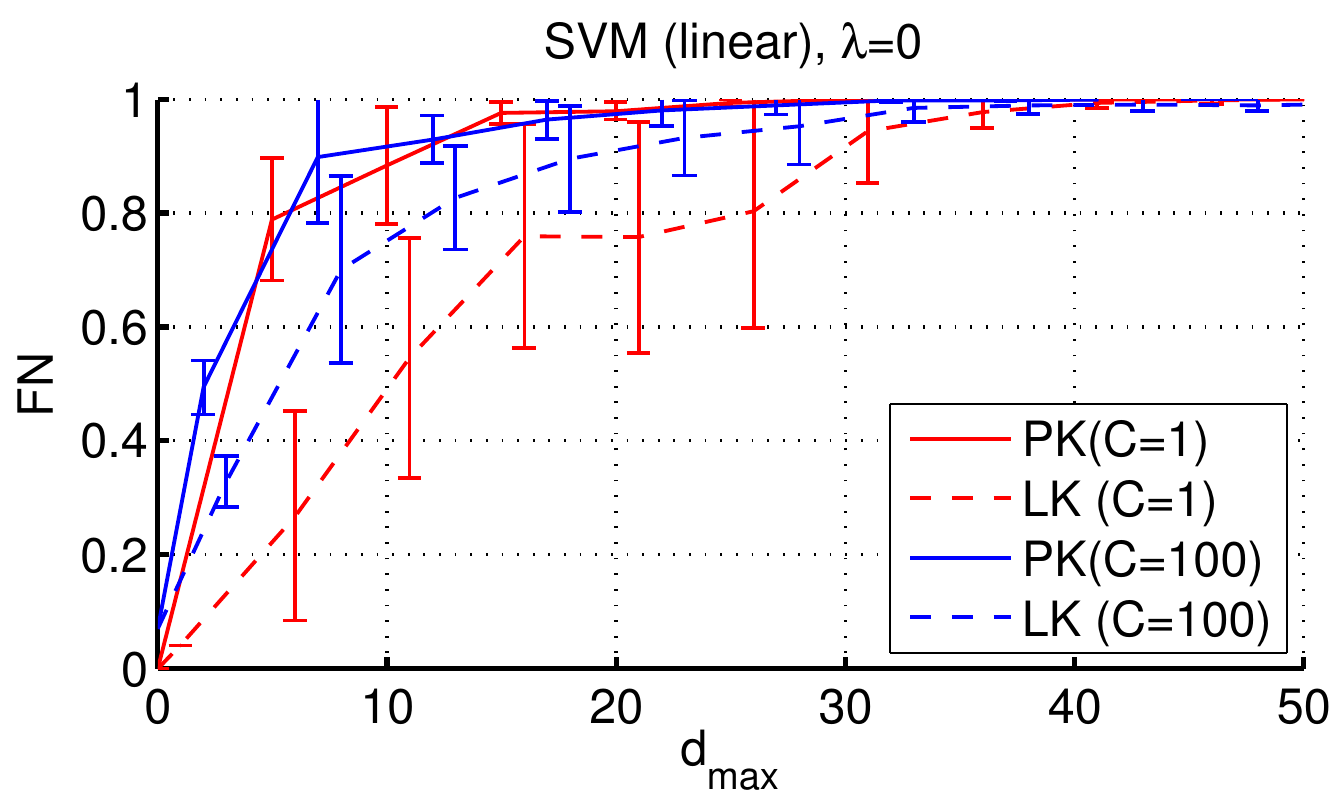}
\includegraphics[width=0.495\textwidth]{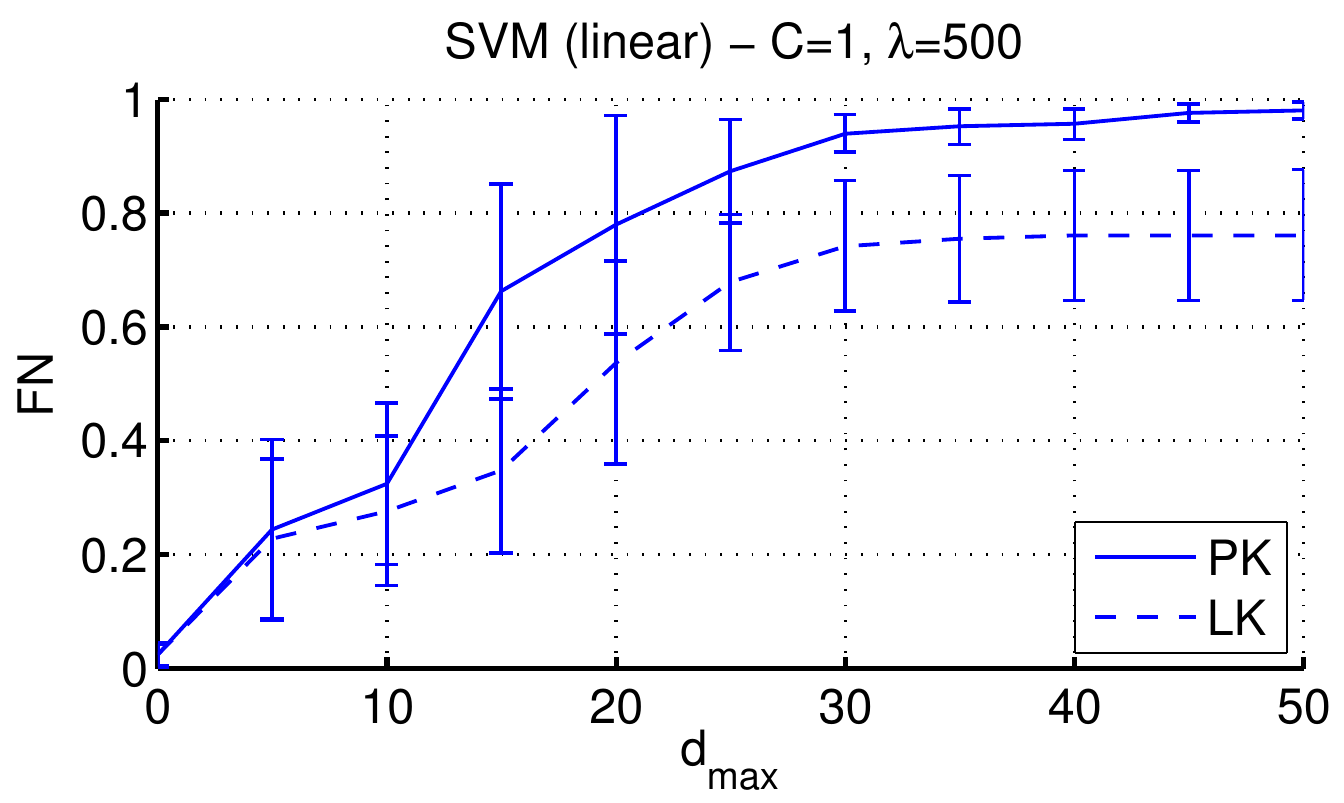}
\includegraphics[width=0.495\textwidth]{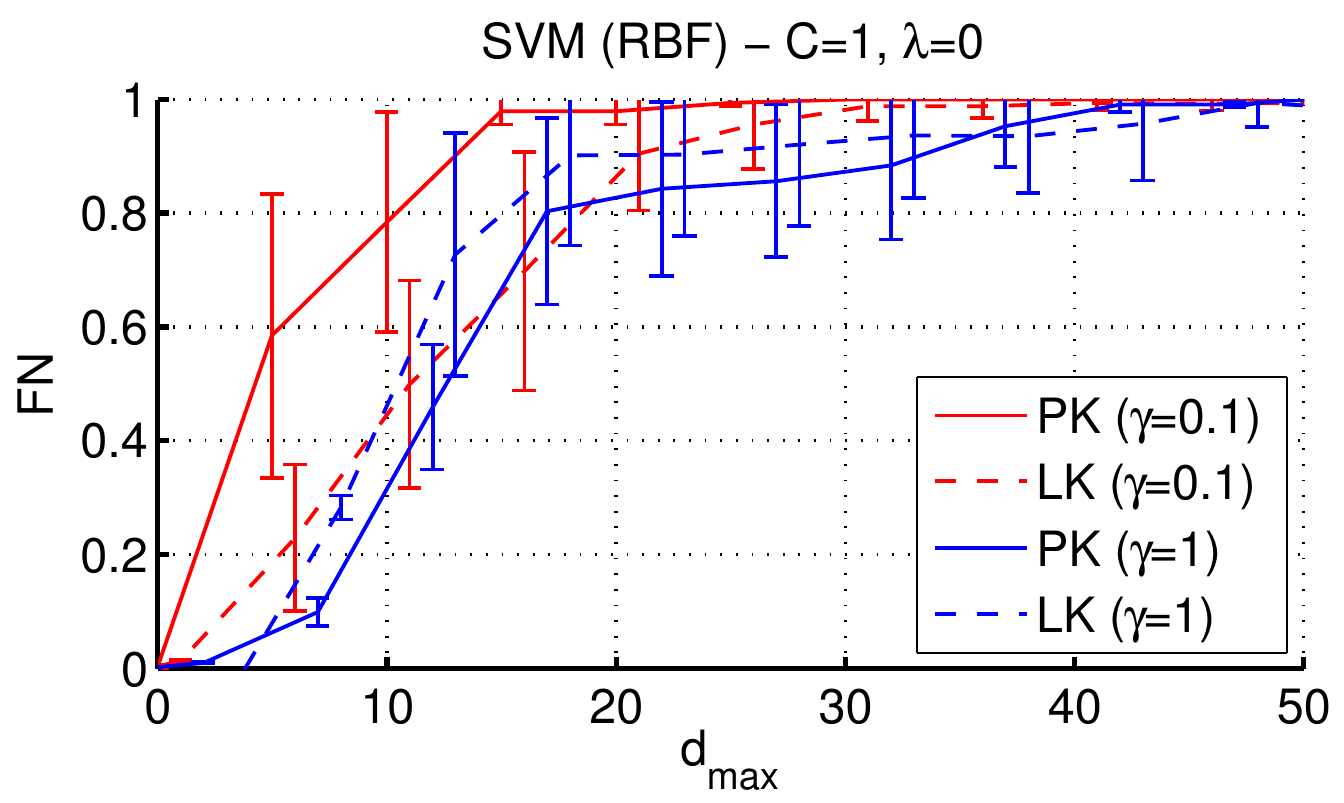}
\includegraphics[width=0.495\textwidth]{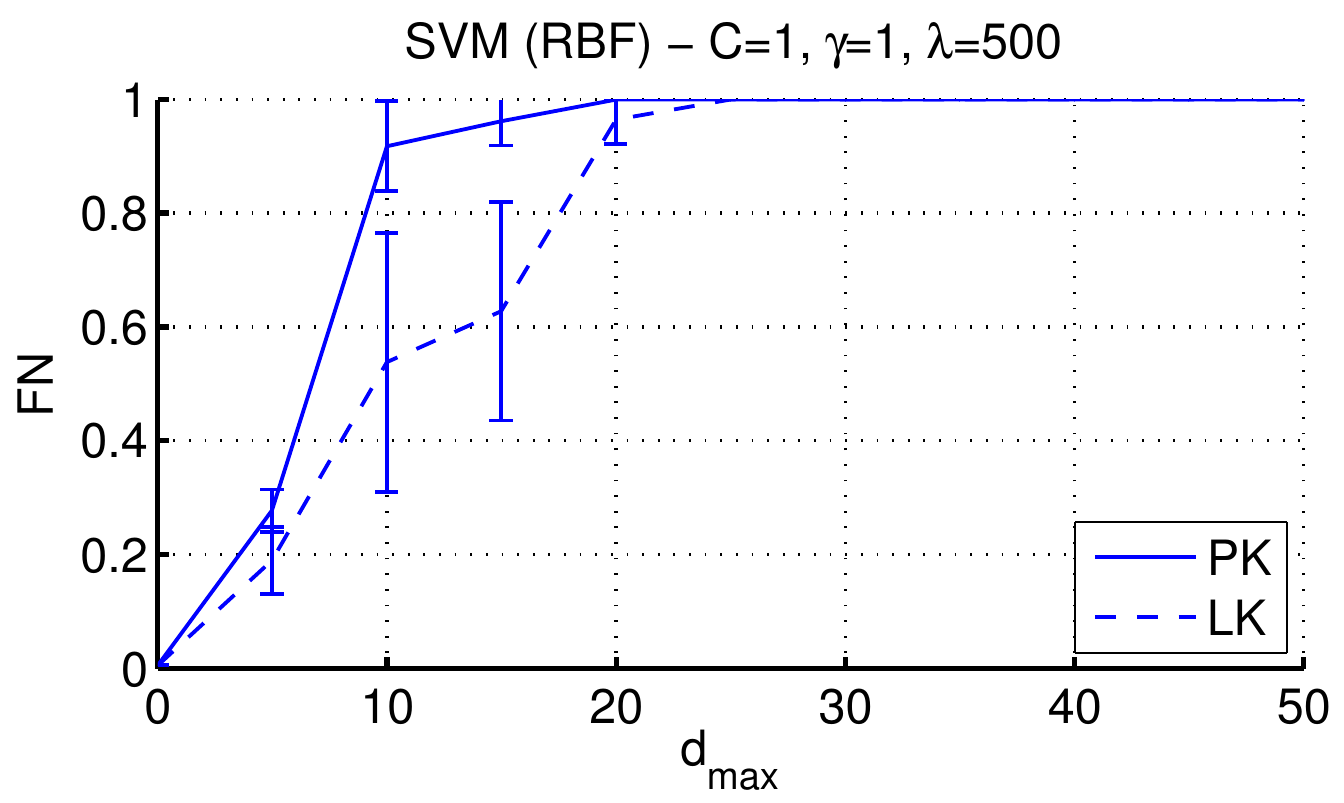}
\includegraphics[width=0.495\textwidth]{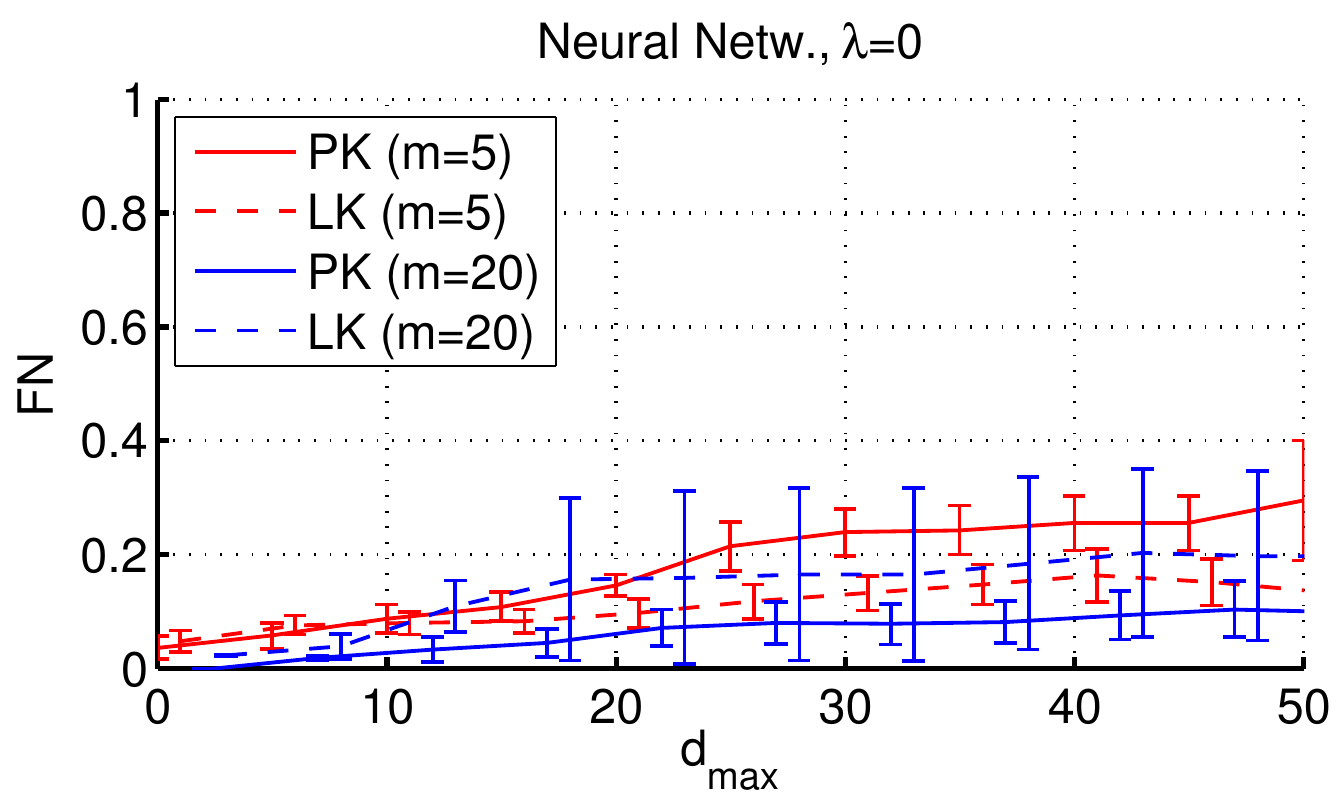}
\includegraphics[width=0.495\textwidth]{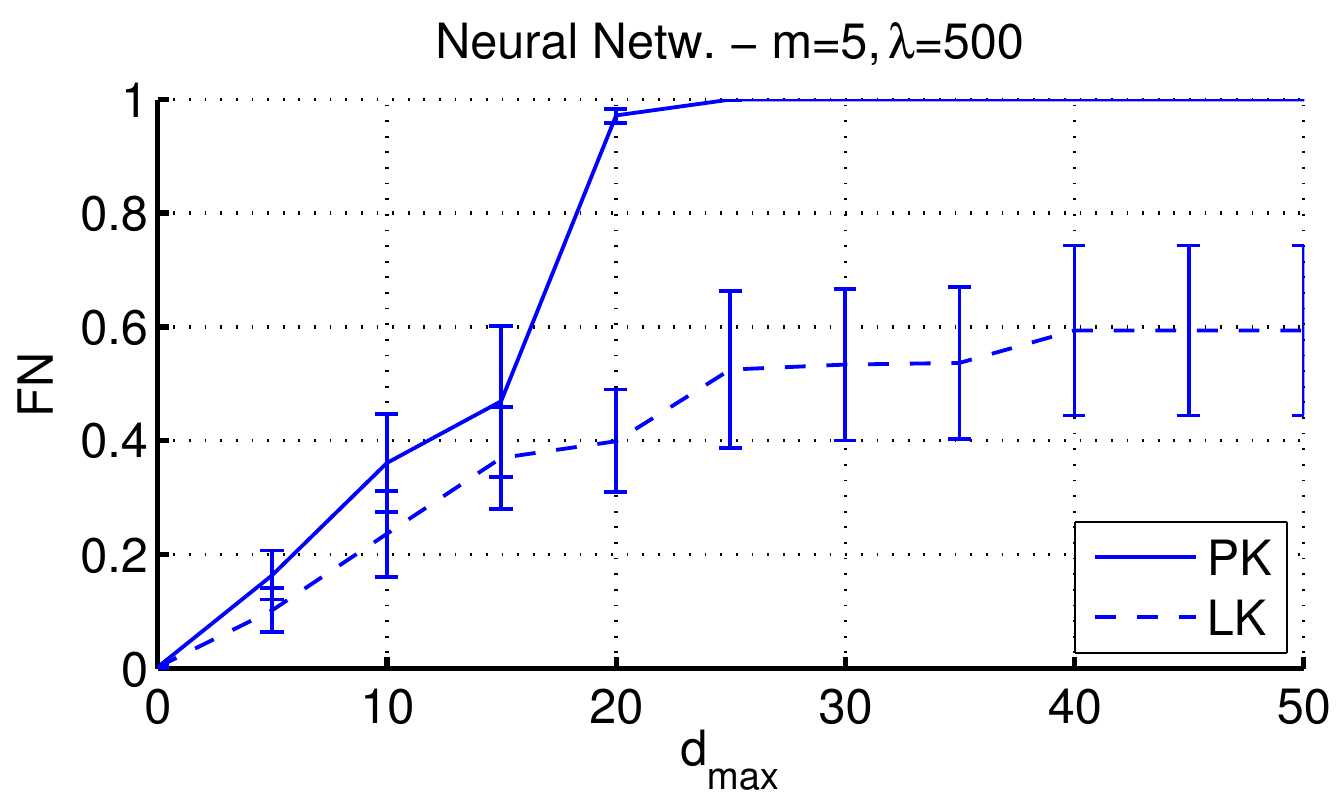}
\caption{Experimental results for SVMs with linear and RBF kernel (first and second row), and for neural networks (third row). We report the FN values (attained at FP=0.5\%) for increasing $d_{\rm max}$. For the sake of readability, we report the average FN value $\pm$ half standard deviation (shown with error bars). Results for perfect (PK) and limited (LK) knowledge attacks with $\lambda=0$ (without mimicry) are shown in the first column, while results with $\lambda=500$ (with mimicry) are shown in the second column. In each plot we considered different values of the classifier parameters, \ie, the regularization parameter $C$ for the linear SVM, the kernel parameter $\gamma$ for the SVM with RBF kernel, and the number of neurons $m$ in the hidden layer for the neural network, as reported in the plot title and legend.} 
\label{fig:results}
\end{center}
\end{figure}

\medskip
\noindent \textbf{Experimental results.} We report our results in Figure~\ref{fig:results}, in terms of the
false negative (FN) rate attained by the targeted classifiers
as a function of the maximum allowable number of modifications, $d_{\rm max} \in [0, 50]$. We compute the FN rate corresponding to a fixed false positive (FP) rate of FP$=0.5\%$. For $d_{\rm max}=0$, the FN rate
corresponds to a standard performance evaluation
using unmodified PDFs.
As expected, the FN rate increases with $d_{\rm max}$ as the PDF is increasingly modified. Accordingly, a more secure classifier will exhibit a more
graceful increase of the FN rate.  

\smallskip
\noindent \textit{Results for $\lambda=0$.} We first investigate the effect of the proposed attack in the PK case,
without considering the mimicry component (Figure~\ref{fig:results},
first column), for varying parameters of the considered classifiers.
The linear SVM (Figure~\ref{fig:results}, top-left plot) is almost
always evaded with as few as $5$ to $10$ modifications, independent of the
regularization parameter $C$.
It is worth noting that attacking a linear classifier
amounts to always incrementing the value of the same highest-weighted feature (corresponding to the \texttt{/Linearized} keyword in the majority of the cases) until it reaches its upper bound.
This continues with the next highest weighted non-bounded feature until termination. This occurs simply because the gradient of $g(\mathbf x)$ does not depend on $\mathbf x$ for a linear classifier (see Section~\ref{sect:gradient-disc}).
With the RBF kernel (Figure~\ref{fig:results}, middle-left plot), SVMs
exhibit a similar behavior with $C=1$ and various values of its
$\gamma$ parameter,\footnote{We also conducted experiments using $C=0.1$ and
  $C=100$, but did not find significant differences compared to the
  presented results using $C=1$.} and the RBF SVM provides a higher degree of security compared
to linear SVMs (\emph{cf.} top-left plot and middle-left plot in
Figure~\ref{fig:results}). Interestingly, compared to SVMs, neural networks  (Figure~\ref{fig:results}, bottom-left plot) seem to be much more robust against the proposed evasion attack. This behavior can be explained by observing that the decision function of neural networks may be characterized by flat regions (\ie, regions where the gradient of $g(\mathbf x)$ is close to zero). Hence, the gradient descent algorithm based solely on $g(\mathbf x)$ essentially stops after few attack iterations for most of the malicious samples, without being able to find a suitable attack.

In the LK case, without mimicry, classifiers are evaded with a probability only \emph{slightly} lower than that found in the PK case, even
when only $n_{g}=100$ surrogate samples are used to learn the surrogate classifier. This aspect highlights the threat posed by a skilled adversary with incomplete knowledge: only a small set of samples may be required to successfully attack the target classifier using the proposed algorithm.

\smallskip
\noindent \textit{Results for $\lambda=500$.} When mimicry is used (Figure~\ref{fig:results}, second column), the success of the evasion of linear SVMs (with $C=1$) decreases both in the PK
 (\eg, compare the blue curve in the top-left plot with the solid blue curve in the top-right plot)
and LK case
 (\eg, compare the dashed red curve in the top-left plot with the dashed blue curve in the top-right plot).
The reason is that the computed direction tends to lead to a slower descent; \ie, a less direct path that often requires more modifications to evade the classifier.
In the non-linear case (Figure~\ref{fig:results}, middle-right and bottom-right plot), instead, mimicking exhibits some beneficial aspects for the attacker, although the constraint on feature addition may make it difficult to properly mimic legitimate samples. In particular, note how the targeted SVMs with RBF kernel (with $C=1$ and $\gamma=1$) in the PK case (\eg, compare the solid blue curve in the middle-left plot with the solid blue curve in the middle-right plot) is evaded with a significantly higher probability than in the case of $\lambda=0$.
The reason is that, as explained at the end of Section~\ref{sect:attack-strategy}, a pure descent strategy on $g(\mathbf x)$ may find local minima (\ie, attack samples) that do not evade detection, while the mimicry component biases the descent towards regions of the feature space more densely populated by legitimate samples, where $g(\mathbf x)$ eventually attains lower values.
For neural networks, this aspect is even more evident, in both the PK and LK settings (compare the dashed/solid curves in the bottom-left plot with those in the bottom-right plot), since $g(\mathbf x)$ is essentially flat far from the decision boundary, and thus pure gradient descent on $g$ can not even commence for many malicious samples, as previously mentioned. In this case, the mimicry term is thus critical for finding a reasonable descent path to evasion.

\medskip
\noindent \textbf{Discussion.} Our attacks raise questions about the feasibility of detecting
malicious PDFs solely based on logical structure. We found that \texttt{/Linearized}, \texttt{/OpenAction}, \texttt{/Comment}, \texttt{/Root} and \texttt{/PageLayout} were among the most commonly manipulated keywords. They indeed are found mainly in
legitimate PDFs, but can be easily added to malicious PDFs by the
versioning mechanism. The attacker can simply insert comments inside
the malicious PDF file to augment its \texttt{/Comment} count.
Similarly, she can embed \emph{legitimate} OpenAction code to add
\texttt{/OpenAction} keywords or  add new pages to insert
\texttt{/PageLayout} keywords.

\section{Conclusions, limitations and future work}
\label{sect:conclusions}

In this work we proposed a simple algorithm for evasion of classifiers
with differentiable discriminant functions. We investigated the attack
effectiveness in the case of perfect and limited knowledge of the
attacked system, and empirically showed that very popular
classification algorithms (in particular, SVMs and neural networks)
can still be evaded with high probability even if the adversary can
only learn a copy of the classifier from a small surrogate dataset.
Thus, our investigation raises important questions on whether such
algorithms can be reliably employed in security-sensitive
applications.

We believe that the proposed attack formulation can be extended to classifiers with non-differentiable discriminant functions as well, such as decision trees and $k$-nearest neighbors; \eg, by defining suitable search heuristics similar to our mimicry term to minimize $g(\mathbf x)$. 

Interestingly our analysis also suggests improvements for classifier
security.  From Fig.~\ref{fig:attack-strategy}, it is clear that a
tighter \emph{enclosure} of the legitimate samples increasingly forces
the adversary to mimic the legitimate class, which may not always be
possible; \eg, malicious network packets or PDF files must contain a
valid exploit for the attack to be successful.
Accordingly, more secure classifiers can be designed by employing
regularization terms that promote
enclosure of the legitimate class; \eg, by penalizing ``blind spots''
- regions with low $p(\mathbf x)$ - classified as legitimate. 
Alternatively, one may explicitly model the attack distribution, as in \cite{biggio11-smc};
or add the generated attack samples to the training set. 
Nevertheless, improving
security probably must be balanced with a higher FP rate.

In our example applications, the feature representations could be \emph{inverted} to obtain a corresponding real-world objects (\eg, spam emails, or PDF files); \ie, it is straightforward to manipulate the given real-world object to obtain the desired feature vector $\mathbf x^{*}$ of the \emph{optimal} attack.
However, in practice some complex feature mappings can not be easily inverted; \eg, $n$-gram features~\cite{fogla06}. 
Another idea would be to modify the real-world object at each step of the gradient descent to obtain a sample in the feature space which is as close as possible to the sample that would be obtained at the next attack iteration. A similar technique has been already exploited by \cite{biggio12-icml} to overcome the pre-image problem.

Other interesting extensions of our work may be to (\emph{i}) consider
more effective strategies such as those proposed by
\cite{lowd05,nelson12-jmlr} to build a small but representative set of
surrogate data; and (\emph{ii}) improve the classifier estimate $\hat
g(\mathbf x)$. To this end, one may exploit ensemble techniques such
as bagging or the random subspace method to train several classifiers
and then average their output.

\subsubsection*{Acknowledgments.} This work has been partly supported by the project CRP-18293, L.R. 7/2007, Bando 2009, and by the project ``Advanced and secure sharing of multimedia data over social networks in the future Internet'' (CUP F71J11000690002), both funded by Regione Autonoma della Sardegna. Davide Maiorca gratefully acknowledges Regione Autonoma della Sardegna for the financial support of his PhD scholarship. Blaine Nelson thanks the Alexander von Humboldt Foundation for providing additional financial support. The opinions expressed in this paper are solely those of the authors and do not necessarily reflect the opinions of any sponsor.

\end{document}